\documentclass[aip,preprint,superscriptaddress]{revtex4-1} 

\usepackage{amssymb,amsmath,graphicx,natbib,hyperref,lineno}

\begin{document}
\linenumbers
\title{Direct observation of dynamic surface acoustic wave controlled carrier injection into single quantum posts using phase-resolved optical spectroscopy}

\author{S. V\"{o}lk}
\affiliation{Lehrstuhl f\"{u}r Experimentalphysik 1 and Augsburg Centre for Innovative Technologies (ACIT), Universit\"{a}t Augsburg, 86159 Augsburg, Germany} 
\author{F. Knall}
\affiliation{Lehrstuhl f\"{u}r Experimentalphysik 1 and Augsburg Centre for Innovative Technologies (ACIT), Universit\"{a}t Augsburg, 86159 Augsburg, Germany} 
\author{F. J. R. Sch\"{u}lein}
\affiliation{Lehrstuhl f\"{u}r Experimentalphysik 1 and Augsburg Centre for Innovative Technologies (ACIT), Universit\"{a}t Augsburg, 86159 Augsburg, Germany} 
\author{T. A. Truong}
\affiliation{Materials Department, University of California, Santa Barbara CA 93106, United States}
\author{H. Kim}
\affiliation{Physics Department, University of California, Santa Barbara CA 93106, United States}
\author{P. M. Petroff}
\affiliation{Materials Department, University of California, Santa Barbara CA 93106, United States}
\author{A. Wixforth}
\affiliation{Lehrstuhl f\"{u}r Experimentalphysik 1 and Augsburg Centre for Innovative Technologies (ACIT), Universit\"{a}t Augsburg, 86159 Augsburg, Germany} 
\author{H. J. Krenner}\email{hubert.krenner@physik.uni-augsburg.de}
\affiliation{Lehrstuhl f\"{u}r Experimentalphysik 1 and Augsburg Centre for Innovative Technologies (ACIT), Universit\"{a}t Augsburg, 86159 Augsburg, Germany}


\begin{abstract}
A versatile stroboscopic technique based on active phase-locking of a surface acoustic wave to picosecond laser pulses is used to monitor dynamic acoustoelectric effects. Time-integrated multi-channel detection is applied to probe the modulation of the emission  of a quantum well for different frequencies of the surface acoustic wave. For quantum posts we resolve dynamically controlled generation of neutral and charged excitons and preferential injection of holes into localized states within the nanostructure. 
\end{abstract}

\maketitle

\begin{figure}[f]
	\begin{center}
		\includegraphics[width=0.95\columnwidth]{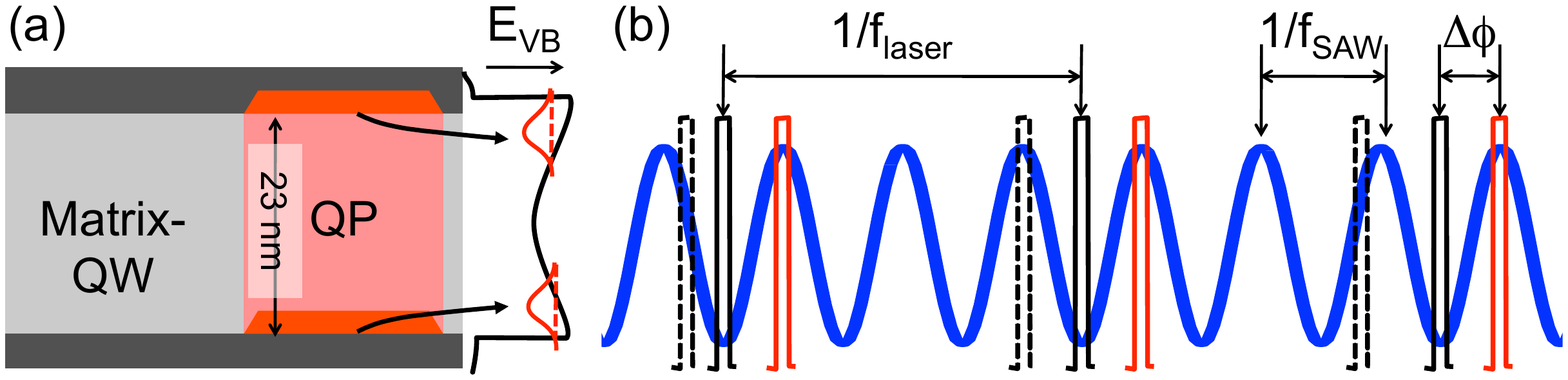}
		\caption{(Color online) (a) Schematic of self-assembled QPs embedded in a lateral matrix QW and localization of holes in local valence band minima at ends of the QP. (b) Scheme to active locking and tuning (solid) of the SAW (bold) phase and the laser pulses (solid) or averaging over all phases (dashed).}
		\label{fig:1}
	\end{center}
\end{figure}

Surface acoustic waves (SAWs) are a powerful tool for the investigation of electrically and optically active low-dimensional semiconductor structures at frequencies exceeding several GHz. For optically active structures, examples range from acoustic charge and spin transport \cite{Rocke:97,Rotter:99a,*Sogawa:01a} to the manipulation of optical resonances in quantum dots (QDs) \cite{Gell:08,Couto:09,*Metcalfe:10} and optical cavities \cite{Lima:05}. Due to the high frequencies involved, these optical experiments require a high temporal resolution. This is typically achieved by using single-channel time-correlated single photon counting techniques or high speed, intensified charge coupled device (iCCD) detectors or streak cameras which have, however, relatively poor sensitivity \cite{Gell:08,Couto:09,*Metcalfe:10,Alsina:01,*Sogawa:09b}.\\
In this letter, we demonstrate \emph{dynamic} acoustically controlled carrier injection into the confined levels of self-assembled quantum posts (QPs)\cite{Voelk:10b} and the dynamic modulation of the photoluminescence (PL) emission of a quantum well (QW) using an easy to implement stroboscopic technique in which a picosecond diode laser is phase-locked to the SAW. Compared to conventional schemes our technique allows for a fully time-integrated multi-channel detection with a phase resolution limited by the radiative lifetime of the emitter system, only. \\
For our experiments, we use a sample containing a single layer of 23 nm high QPs grown by molecular beam epitaxy \cite{He:07,*Li:08c}. These QPs are columnar nanostructures embedded in a lateral matrix QW as shown in Fig. \ref{fig:1} (a) in which hole states are localized at the two ends as indicated in the sketched valence band. Details on the applied growth technique using a stacking sequence, the basic optical properties and device applications of QPs are described elsewhere \cite{Krenner:09,*Ridha:10}. For device processing we select a region with a sufficiently low surface density $(<1~\mathrm{\mu m^{-2}})$ of QPs where we can probe (i) individual QPs and (ii) the lateral matrix QW surrounding the QPs.  On this sample a set of lithographically defined interdigital transducers (IDTs) allow for SAW excitation at frequencies of $f_1=232.2$ MHz, $f_2=538.3$ MHz and $f_3=968.5$ MHz at low temperatures, respectively. 

We study the emission of these QPs and the lateral QW\cite{He:07,*Li:08c} by conventional low temperature ($T\sim10$ K) $\mathrm{\mu}$-PL. Carriers are photogenerated in the GaAs bulk using a diode laser emitting $\tau_{laser}\lesssim 100 \rm ~ps$ long pulses at $\lambda = 661$ nm focused to a $\sim3~\mathrm{\mu m}$ spot using a $50\times$ microscope objective. The signal is collected via the same lens and detected using a 0.5 m imaging grating monochromator and a highly sensitive, liquid nitrogen cooled, Si CCD camera. The excitation laser can be triggered either by an internal clock or using an external signal used for synchronization with the RF signal applied to the IDTs to generate a SAW. In the experiments presented here, the SAW is applied in pulses (on/off duty cycle 1:9, $f_{rep}\sim100~\mathrm{kHz}$). As shown in Fig. \ref{fig:1} (b) the excitation laser pulses can be phase-locked for the SAW frequency being an harmonic of the laser repetition rate: $f_{\rm SAW} = n\cdot f_{\rm laser}$. Under these conditions the short laser pulse is exciting the sample at a constant phase $\phi$ i.e. fixed time of the SAW oscillation. The full SAW cycle can then be resolved by variation of the phase offset $\Delta\phi$ between the SAW and the train of laser pulses. This is shown for $\Delta\phi=180\mathrm{^o}$ for the solid black and red (gray) laser pulses in Fig. \ref{fig:1} (b). Thus, for a sufficiently fast radiative process with a PL decay time constant $\tau_{\rm PL}\cdot f_{\rm SAW}< 1/2$, we are able to obtain \emph{full} phase information. Moreover, simply by choosing an arbitrary phase relation [dashed laser pulses in Fig. \ref{fig:1} (b)], phase-averaged spectra are recorded.\\
\begin{figure}[f]
	\begin{center}
		\includegraphics[width=0.65\columnwidth]{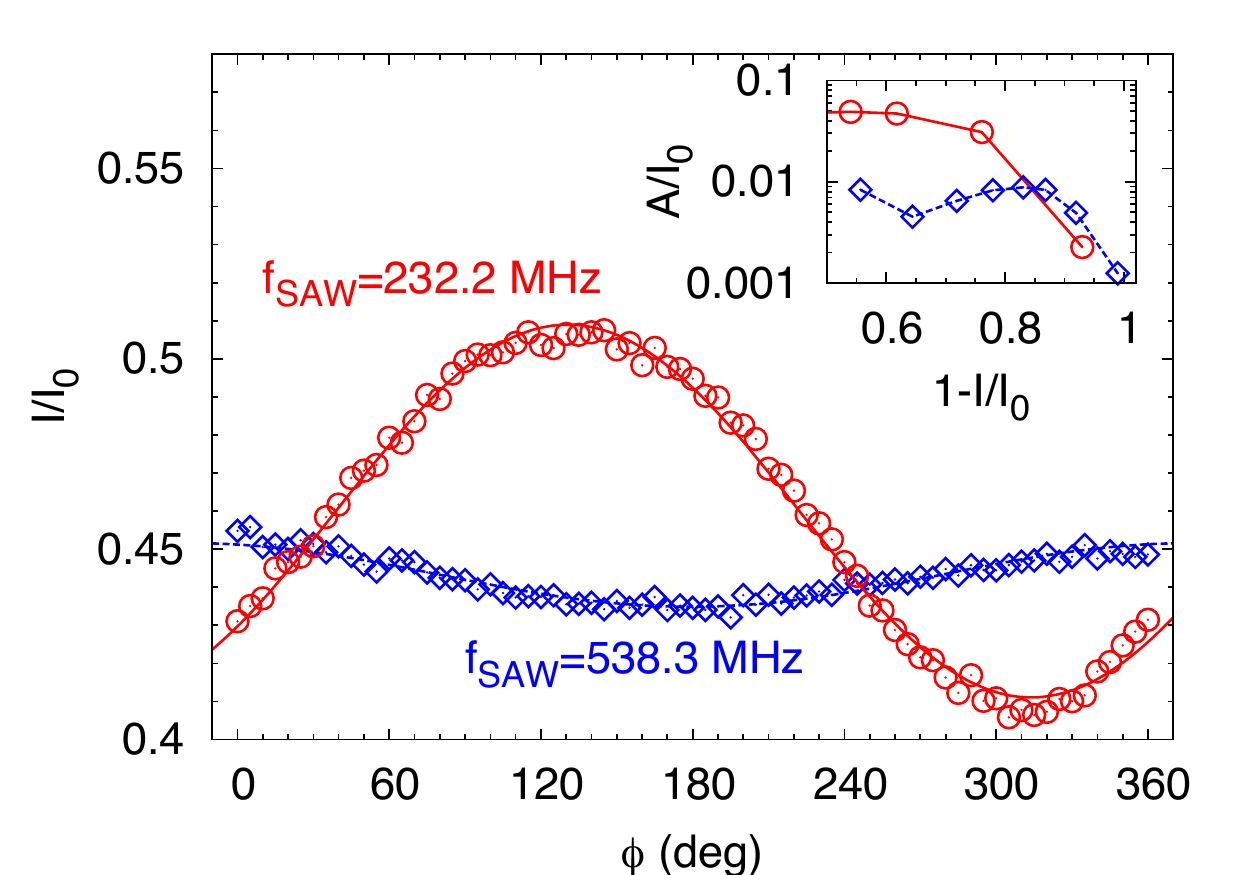}
		\caption{(Color online) Main panel: Integrated and normalized QW emission (symbols) and sine fits (lines) for SAW frequencies $f_1$ (circles) and $f_2$ (diamonds)as a function of the local phase $\phi$. Inset: Relative amplitude of the oscillation for different levels of SAW induced emission quenching $(1-I/I_0)$. }
		\label{fig:2}
	\end{center}
\end{figure}
To study the phase and thus temporal resolution of this technique, we start by investigating the dynamic quenching of the matrix QW PL for different $ f_{\rm SAW}$ \cite{Voelk:10b}. We actively lock the laser pulses to the SAW by setting $n=3,~7,~13$ for $f_1$, $f_2$ and $f_3$, respectively. To achieve comparable experimental conditions, we applied RF powers for which a similar suppression of the PL was observed for these frequencies. As we tune $\phi,$ [c.f. Fig. \ref{fig:2}] we observe a clear oscillatory behavior of the normalized QW emission intensity ($I/I_0$, symbols) arising from the dynamic modulation of band edges which are well reproduced by fitting sine functions with the period of the SAW (lines) and agrees well with previous work\cite{Alsina:01}. The absolute value of $\phi$ depends on the distance of the laser spot to the IDT which leads to the finite offset for the two frequencies and the data on QPs presented later. The amplitude $(A/I_0)$ of this oscillation is significantly reduced for the higher frequency since the time between the maximum and minimum SAW amplitude given by $1/(2\cdot f_2) =0.93\mathrm{~ns}$ is shorter than the radiative decay time of the QW $\tau_{\rm PL}=1.25{\rm ~ns}$. This is further confirmed by a detailed analysis of $A/I_0$ as a function of the suppression of the QW PL $(1-I/I_0)$ for $f_1$ and $f_2$ shown as an inset of Fig. \ref{fig:2}. For the highest frequency $(f_3)$ no more modulation is observed (data not shown). Thus, our stroboscopic technique can be readily applied to resolve periodically driven processes for radiative processes with $\tau_{\rm PL}\cdot f_{\rm SAW} \lesssim1/2$.\\
\begin{figure}[f]
	\begin{center}
		\includegraphics[width=0.9\columnwidth]{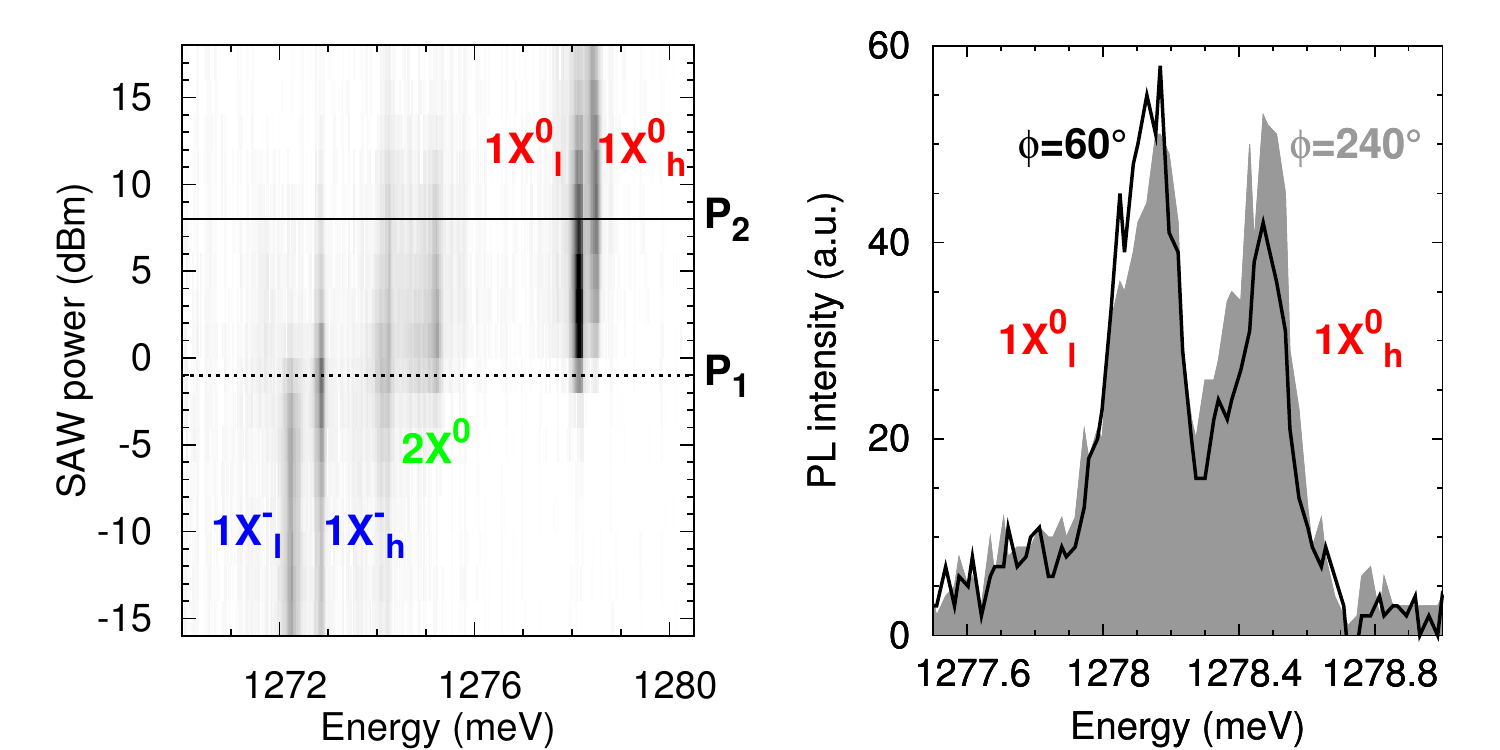}
		\caption{(Color online) (a) Phase-averaged, RF-power dependent PL of a single QP for $f_1$ plotted in grayscale representation. Switching between $(1X^-_{h,l})$ and $(1X^0_{h,l})$ occurs at $P_1 = -1{\rm ~dBm}$. (b) Emission of $1X_{h,l}^0$ at $P_2$ for $\phi=60^{\rm o}$ and $240^{\rm o}$ showing a pronounced modulation of the doublet.}
		\label{fig:3}
	\end{center}
\end{figure}
We also applied the technique to QPs to monitor the dynamic sequential carrier injection controlled by a  SAW $(f_1=232.2\mathrm{~MHz})$. These nanostructures are particularly suited due to the efficient acoustic charge conveyance in the lateral matrix QW. In Fig. \ref{fig:3} (a) we present SAW phase-averaged PL spectra recorded under weak optical pumping from an individual QP as a function of the applied RF power (grayscale representation). We observe a characteristic switching\cite{Voelk:10b} from an emission doublet arising from recombination of the negatively charged exciton $(1X^-_{l,h})$ to a doublet at higher energies assigned to the charge neutral single exciton $(1X^0_{l,h})$. The indices $l~(h)$ denote the lower (higher) energy line of the doublet. These doublets arise from recombinations between \emph{delocalized} electrons and the two non-degenerate hole levels\cite{Krenner:08b} at the two ends of the QP [c.f. Fig. \ref{fig:1} (a)]. We performed full SAW phase scans at the two characteristic power levels marked in Fig. \ref{fig:3} (a): $P_{1} = -1{\rm ~dBm}$ is close to the onset of charge conveyance and the emission spectra is dominated by emission of $1X^-_{l,h}$. At $P_{2} = +8{\rm ~dBm},$ efficient charge conveyance occurs, the $1X^-_{l,h}$ PL lines are strongly suppressed and emission of $1X^0_{l,h}$ is detected. Under the applied experimental conditions the band gap modulation induced by the SAW \cite{Alsina:01} is not sufficient to \emph{spectrally tune} individual PL lines \cite{Gell:08,Voelk:10b}. In Fig. \ref{fig:3} (b) we compare the $1X^0_{l,h}$ doublet at $P_2$ for two distinct relative SAW phases split by $\Delta \phi = 180^{\rm o}$. For $\phi = 60^{\rm o}$ (gray shaded) the lower energy line $1X^0_{l}$ of the doublet is more intense than $1X^0_{h}$ whilst for $\phi = 240^{\rm o}$ (line) both lines have approximately the same intensity. Since the two lines of the doublet arise from localization of holes at either the upper or the lower end of the QP, this observation demonstrates that the injection of this carrier species into the QPs also depends on the local $\phi$.\\

\begin{figure}[f]
	\begin{center}
		\includegraphics[width=0.9\columnwidth]{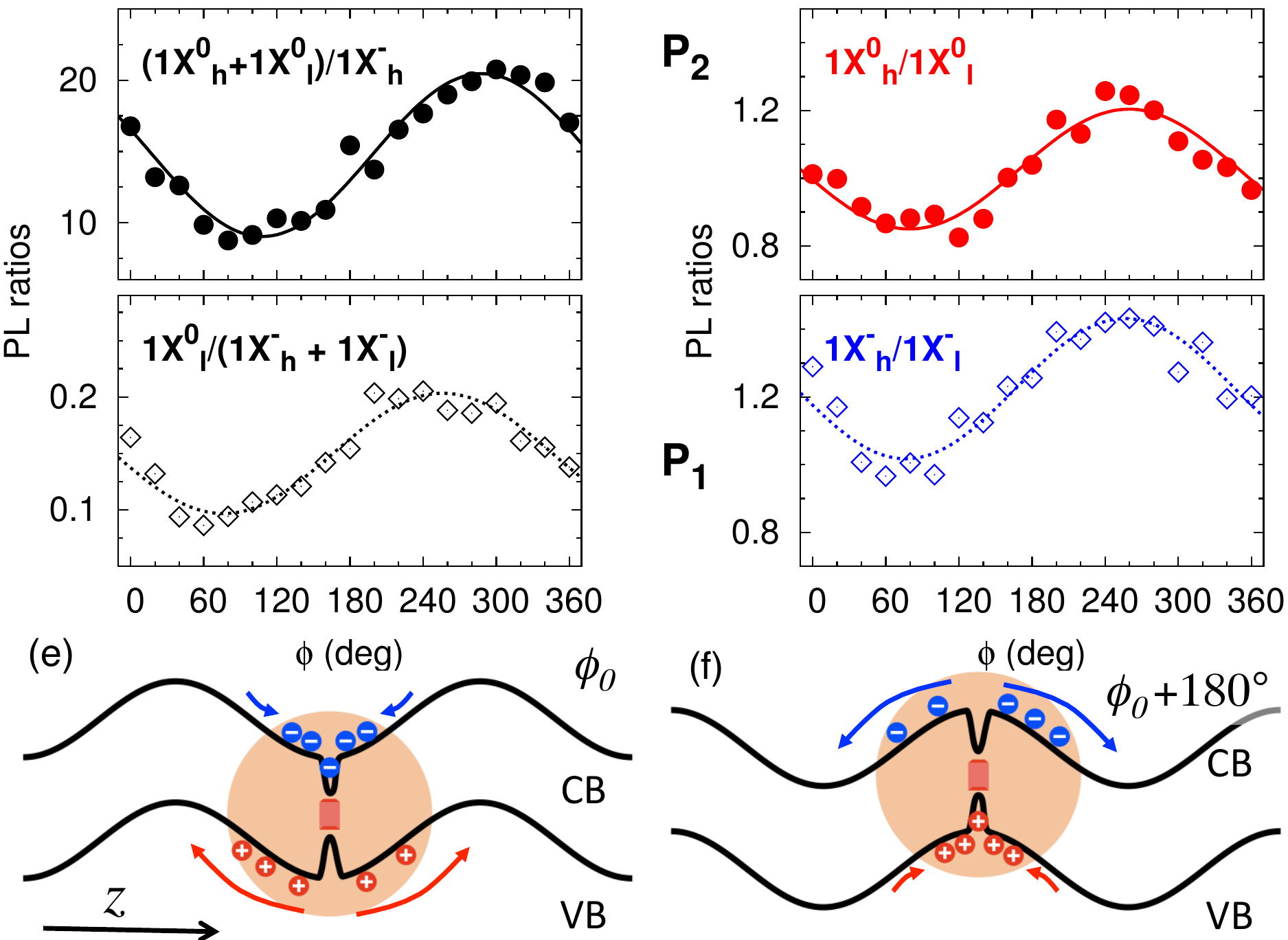}
		\caption{(Color online) Relative intensities (symbols) of $1X^0$ and $1X^-$ (a) and (c) and  $1X_{h,l}$ doublets (b) and (d) for a $f_1$ at $P_1$ (lower panels) and $P_2$ (upper panels) as a function of $\phi$. The corresponding fits are plotted as lines. (e) and (f)  SAW-induced band edge modulation: At $\phi_0 = 80^{\rm o}$ and $\phi_0+180^{\rm o}$ attractive potentials for electrons and holes are locally induced by the SAW at the QP position.}
		\label{fig:4}
	\end{center}
\end{figure}
The detailed investigation of the SAW controlled carrier capture and a complete analysis of the relative PL intensities of different excitonic species and emission doublets is presented in Fig. \ref{fig:4}. We plot the relative intensities of $1X^0$ and $1X^-$, $I_{1X^0}/I_{1X^-}$  as a function of the local SAW phase in Fig. \ref{fig:4} (a) and (c) and the relative intensities of the two lines forming the emission doublets, $I_{1X_h}/I_{1X_l}$, in Fig. \ref{fig:4} (b) and (d) at both power levels ($P_1$ diamonds, $P_2$ circles). Clearly, all ratios show the same oscillatory behavior as a function of $\phi,$ providing direct evidence for dynamic carrier capture directly controlled by the SAW. In particular, within this range of SAW powers the generation of charged and neutral excitons is highly dependent on $\phi$. This follows from our observation that the relative intensity detected from the $1X^0_{l,h}$ is reduced by a factor of 2 at $\phi_0=80^{\rm o}$ compared to $\phi_0+180^{\rm o}$ [c.f. Fig. \ref{fig:4} (a)]. As can be seen from both experimental data (symbols) and the fits (lines), at both power levels [c.f. Fig. \ref{fig:4} (a) and (f)] the maxima occur at almost identical $\phi$ indicating the same driving mechanism in both cases. The SAW induces an Type-II band edge modulation giving rise to local accumulation of electrons or holes\cite{Alsina:02a} at distinct values of $\phi_{\rm SAW}$ as shown schematically in Fig. \ref{fig:4} (e) and (f). At $\phi_0$ an attractive minimum is formed for electrons in the conduction band while the situation is inverted at $\phi_0+180^{\rm o}$ and holes are accumulated at a stable maximum in the valence band. Thus, the local concentration of the two carrier species is dynamically controlled locally by $\phi$ leading to the observed modulation of the QP occupancy. From the relative intensity of $I_{1X^0}/I_{1X^-}$  we conclude that (i) $1X^-$ is predominantly formed around $\phi_0$ where electrons are collected at the position of the QP whilst (ii) a local accumulation of holes is required to form $1X^0$.  The finite phase-shift between $P_1$ and $P_2$ [c.f. Fig \ref{fig:4} (a) and (c)] could arise from contributions between lateral and vertical piezoelectric fields.
Furthermore, we investigated in detail the modulation with the emission doublet $1X_{h,l}$. For both $1X^0$ and $1X^-,$ the same phase dependence of the ratio of the energetically higher $1X_{h}$ and the lower $1X_{l}$ line is observed in the experimental data (symbols) with the corresponding fits (lines) in Fig. \ref{fig:4} (b) and (d). Moreover, the intensity of $1X_{h}$ is increased by about a factor of 2 at $\phi_0+180^{\rm o}$ compared to $\phi_0$. This finding indicates that at these distinct phases the hole is injected predominantly at one specific end of the QP. The preferential injection could be driven by the vertical electric field component at these local SAW phases. This finite electric field gives rise to a enhanced hole localization at one side of the lateral QW leading to an increase in the population of the corresponding localized hole level in the QP. This interpretation is in good agreement with the observed localization (delocalization) of holes (electrons) within the QP for weak static vertical electric fields \cite{Krenner:08b,Krenner:08a}. Since the oscillation of $1X_{h}$ and $1X_{l}$ are in phase for both exciton species we conclude that the major contribution to the energetic splitting within the doublet indeed arises from the distinct single particle energies of the localized hole states at the two ends of the QP.\\

In summary, we demonstrated an stroboscopic technique to probe the influence of a SAW on optical emitters using time-integrated multi-channel detection based on active phase-locking of a SAW to the excitation laser pulses. It can be applied to monitor the dynamic modulation of the emission of a QW and the dynamic SAW controlled injection of carriers into the confined levels of QPs. Our versatile technique is limited only by the radiative lifetime of the studied system and can be readily applied to investigate a wide range of optically active nanosystems such as different types of single and coupled QDs \cite{Gell:08,Couto:09,*Metcalfe:10,Schuelein:09,*Krenner:05a}, microcavities\cite{Lima:05,Stoltz:05,*Kress:05} or nanostructures directly transferred onto piezoelectric substrates\cite{Talyanskii:01,*Kouwen:10a,*Fuhrmann:10a}.
\\

This work was supported by the {\it Nanosystems Initiative Munich} (NIM), NSF via NIRT (CCF- 0507295), DARPA SEEDLING and the Humboldt-Foundation.

\end{document}